\documentclass[12pt]{iopart}

\usepackage{iopams}
\usepackage{graphicx}

\begin{document}

\title{Analytical energy spectrum for hybrid mechanical systems}

\author{Honghua Zhong$^{1,2}$, Qiongtao Xie$^{1,3,8}$, Xiwen Guan$^{4,5}$, Murray T. Batchelor$^{6,5,7}$, Kelin Gao$^{4}$, and Chaohong Lee$^{1,8}$}

\address{$^{1}$State Key Laboratory of Optoelectronic Materials and Technologies, School of Physics and Engineering, Sun Yat-Sen University, Guangzhou 510275, China}

\address{$^{2}$Department of Physics, Jishou University, Jishou 416000, China}

\address{$^{3}$School of Physics and Electronic Engineering, Hainan Normal University, Haikou 571158, China}

\address{$^{4}$State Key Laboratory of Magnetic Resonance and Atomic and Molecular Physics, Wuhan Institute of Physics and Mathematics, Chinese Academy of Sciences, Wuhan 430071, China}

\address{$^{5}$Department of Theoretical Physics, Research School of Physics and Engineering, Australian National University, Canberra, ACT 0200, Australia}

\address{$^{6}$Centre for Modern Physics, Chongqing University, Chongqing 400044, China}

\address{$^{7}$Mathematical Sciences Institute, Australian National University, Canberra ACT 0200, Australia}

\address{$^{8}$Authors to whom any correspondence should be addressed.}

\ead{xieqiongtao@gmail.com, chleecn@gmail.com}

\begin{abstract}
We investigate the energy spectrum for hybrid mechanical systems described by non-parity-symmetric quantum Rabi models. 
A set of analytical solutions in terms of the confluent Heun functions and their analytical energy spectrum are obtained. 
The analytical energy spectrum includes regular and exceptional parts, which are both confirmed by direct numerical simulation. 
The regular part is determined by the zeros of the Wronskian for a pair of analytical solutions. 
The exceptional part is relevant to the isolated exact solutions and its energy eigenvalues are 
obtained by analyzing the truncation conditions for the confluent Heun functions. 
By analyzing the energy eigenvalues for exceptional points, we obtain the analytical conditions for the 
energy-level-crossings, which correspond to two-fold energy degeneracy.
\end{abstract}


\maketitle

\section{Introduction}

In recent years, hybrid mechanical systems in which a mechanical oscillator is coupled to a 
microscopic quantum system, such as trapped atoms or ions, solid-state spin qubits, or superconducting devices, 
have stimulated great interest (see, e.g.,~\cite{Treutlein} for a recent review). 
Typical hybrid mechanical systems include quantum dots~\cite{Knobel, Zippilli, Bennett}, 
spin systems~\cite{Rugar, Xue}, superconducting circuits~\cite{Armour, Irish, Martin, Xueb} 
and atom-molecular systems~\cite{Wang, Hunger, Camerer, LeeNJP, LeeSLZT}. 
Hybrid mechanical systems provide new approaches to controlling mechanical objects at the level of quantum mechanics. 
Beyond their fundamental interest, hybrid mechanical systems have promising applications in precision sensing, 
quantum engineering and quantum information processing~\cite{Treutlein}.

In most of the realistic hybrid mechanical systems, the dynamics of the
microscopic quantum system can be treated as a two-level system, and
the vibration of the mechanical oscillator acts as a bosonic mode. 
However, in contrast to a system composed of an independent mechanical oscillator 
and an independent two-level system, in a hybrid mechanical system the motion of the
two-level system and the mechanical oscillator is coupled together. 
It is thus necessary to obtain the energy spectra of the coupled system, 
for which we look to the physics of the quantum Rabi model~\cite{Rabi}.

In quantum optics, the quantum Rabi model describes the simplest interaction between
matter and light. There has been considerable recent progress in the investigation of this model. 
In 2011 Braak found an analytical solution and derived the conditions for determining the energy
spectrum~\cite{Braak, Braakb}. Since then, various aspects of the
quantum Rabi model have been discussed from the theoretical 
viewpoint~\cite{Wolf,Wolfb, Hirokawa, Chen,Yu, Ziegler, Moroz,Maciejewski, Zhong}. 
Many generalized quantum Rabi models have been
studied, including a multi-photon Rabi model~\cite{Travenec},
two-mode Rabi model~\cite{Zhang}, multi-atom Rabi
model~\cite{Chilingaryan},  an $N$-state Rabi
model~\cite{Albert2012}, and the asymmetric quantum Rabi model with
different coupling strengths for the counter-rotating wave and
rotating wave interactions~\cite{Shen, Tomka}.

We have developed a method to
obtain a set of analytical solutions for the simple quantum Rabi
model and have given the analytical conditions for determining its energy
spectrum~\cite{Zhong}. 
In the present paper, we apply this method to obtain the
analytical energy spectrum for hybrid mechanical systems described
by generalized quantum Rabi models. A set of analytical solutions
for the hybrid mechanical systems are given in terms of the
confluent Heun functions. The analytical conditions for determining
the energy spectrum are derived. It is shown that the zeros of the
Wronskian of a pair of analytical solutions give the energy spectrum
except for some specific situations. The  exceptional parts of the
energy spectrum are associated with  isolated exact results under
certain special conditions.  The explicit expressions  for  the
exceptional parts are obtained from the  conditions under which the
confluent Heun functions are terminated as  a finite series.
Additionally, from these isolated exact results, we analytically obtain the
conditions for the occurrence of  the level crossings of the energy
spectrum. 
Our analytical energy spectra are confirmed by direct numerical diagonalization.

\section{Hamiltonian}

We consider a hybrid mechanical system in which the motion of the
microscopic quantum system involves only two states
$\left|\downarrow\right\rangle$ and $\left|\uparrow\right\rangle$
far from other states. 
This two-level microscopic quantum system is coupled to a mechanical oscillator. 
In different physical setups, the coupling mechanisms are different. 
In most situations, the coupling is realized by spatial-dependent electrostatic and/or magnetic fields. 
Therefore, the energy separation $E_{\downarrow\uparrow}$ between the two states
$\left|\downarrow\right\rangle$ and $\left|\uparrow\right\rangle$
depends strongly on the external fields. 
For example, considering a two-level microscopic quantum system coupled to an oscillating
electrode or a vibrating magnetic tip with a small displacement
$\hat{x}$, the energy separation $E_{\downarrow\uparrow}$ may be expanded as
\begin{equation}
E_{\downarrow\uparrow}(\hat{x}) = E_{\downarrow\uparrow}^0 
+\frac{\partial E_{\downarrow\uparrow}(\hat{x})}{\partial \hat{x}} \hat{x} 
+\frac{1}{2}\frac{\partial^2 E_{\downarrow\uparrow}(\hat{x})}{\partial \hat{x}^2} \hat{x}^2 +\ldots
\end{equation}
with $E_{\downarrow\uparrow}^0$ being the energy separation at $\hat{x}=0$. 
Usually, in the above series, the higher order terms beyond the linear term are very small and their effects can be ignored~\cite{Treutlein}. 
As for a harmonic oscillator, one can express the displacement $\hat{x}=x_{0}(a^{\dagger}+a)$ 
with the zero-point displacement $x_{0}=\sqrt{\hbar/(2\omega m^*)}$ and a pair of bosonic operators $(a, a^{\dagger})$, 
where $m^*$ denotes the effective mass. 
Therefore, in units of $\hbar=1$, the hybrid mechanical systems obey the Hamiltonian
\begin{equation}
H=H_0(t)+\omega a^{\dagger}a+g\sigma_z(a^{\dagger}+a).
\end{equation}
Here $H_0$ is the unperturbed Hamiltonian for the two-level system, 
$\omega$ is the oscillating frequency, $\sigma_{z}=\left|\uparrow\rangle\langle\uparrow\right| - \left|\downarrow\rangle\langle\downarrow\right|$ 
is the Pauli matrix for the two-level system, and $g=\frac12 {x_0} \frac{\partial E_{\downarrow\uparrow}(\hat{x})}{\partial \hat{x}}$ 
is the coupling strength between the two-level system and the bosonic mode.

In the hybrid mechanical system of a Cooper Pair Box (CPB) coupled to a gate electrode via the gate capacitance $C_g$, 
the CPB works as a two-level quantum system and it is a small superconducting island of Cooper pairs which 
are coherently coupled to a large reservoir via a Josephson tunnel junction. 
Introducing the number operator $\hat{N}$ for excess Cooper pairs on the island and the phase operator $\hat{\phi}$ 
for the relative phase across the Josephson junction, the CPB can be described by the Hamiltonian
\begin{equation}
H_{CPB}=E_C(\hat{N}-N_g)^2 - E_J \cos(\hat{\phi}), 
\end{equation}
with $E_C$ the charging energy, $E_J$ the Josephson energy, and the dimensionless gate charge $N_g=C_g V_g/(2e)$, 
which can be adjusted by the voltage $V_g$ applied across the gate capacitance $C_g$. 
In experiments, the dynamics of the CPB can be restricted to the two energetically lowest charge states~\cite{charge-qubit}, 
such as $\left|\downarrow\right\rangle = \left|N=n\right\rangle$ and $\left|\uparrow\right\rangle = \left|N=n+1\right\rangle$ 
for $N_g=n+\frac{1}{2}+\Delta_{N_g}$ with $\Delta_{N_g} <1$. 
Thus the CPB can be regarded as a spin-$\frac{1}{2}$ particle in an external magnetic field $\vec{B}$, with 
\begin{equation}
H_0=-\frac{1}{2}\vec{B} \cdot \vec{\sigma}=-\frac12 {E_J} \sigma_x + E_C \Delta_{N_g} \sigma_z, 
\end{equation}
where the effective magnetic field $\vec{B}=(B_x, B_y, B_z)=(E_J, 0, -2 E_C \Delta_{N_g})$ and the Pauli matrices 
$\vec{\sigma}=(\sigma_x, \sigma_y, \sigma_z)$. 
This means that the transverse field $B_x$ is just the Josephson tunneling strength and the longitudinal field $B_z$ 
is determined by the energy separation $E_{\downarrow\uparrow}=E_C \Delta_{N_g}$.

To introduce the coupling between the CPB and the mechanical oscillation,
 according to $E_{\downarrow\uparrow}=E_C \Delta_{N_g}$ and
  $N_g=C_g V_g/(2e)=n+\frac{1}{2}+\Delta_{N_g}$,
one may introduce a spatial-dependent energy separation
$E_{\downarrow\uparrow} (x)$ via a
spatial-dependent capacitance $C_g(x)$.
Replacing the gate electrode with a vibrating mechanical beam,
the capacitance appears as $C_g(x)\approx C_g^0 (1-x/d)$
with the beam displacement $x$ and the gate separation $d$ and so
$\Delta_{N_g} = \Delta_{N_g}^0 - C_g^0 V_g x/(2ed)$.
Expressing the displacement $x$ in terms of the bosonic operators
$(a, a^{\dagger})$, this hybrid mechanical system is seen to be equivalent to 
a generalized quantum Rabi model~\cite{Armour, Irish}
\begin{equation}
H=\omega a^{\dagger}a+g\sigma_z(a^{\dagger}+a)-\Delta
\sigma_x+\epsilon\sigma_z,\label{ha1}
\end{equation}
with $\Delta=E_J/2$, $\epsilon=E_C \Delta_{N_g}^0$,
 the mechanical oscillation frequency $\omega$, and
 the coupling strength
  $g=\frac12 {x_0} \frac{\partial E_{\downarrow\uparrow}(x)}{\partial x} = -\frac14 \frac{x_0}{ed}C_g^0V_g$,
  where $x_{0}$ is  the zero-point displacement uncertainty of the mechanical gate electrode. 
By performing a $\pi/2$ rotation  around the $y$ axis, the above Hamiltonian reads
\begin{equation}
H_R^\epsilon=\omega a^{\dagger}a+g\sigma_x(a^{\dagger}+a)+\Delta
\sigma_z+\epsilon\sigma_x. \label{ha2}
\end{equation}
Here the parity symmetry which couples the mechanical oscillator and the two-level
 system is broken by the term $\epsilon\sigma_x$.
We will thus concentrate our discussion on this non-parity-symmetric quantum Rabi model.

\section{Analytical energy spectrum}

In this section, we present analytical results for the energy spectrum of Hamiltonian~(\ref{ha2}).
We do this by generalizing our recent method for the parity-symmetric quantum Rabi model, for which $\epsilon=0$~\cite{Zhong}.
The analytical solutions for the non-parity-symmetric Hamiltonian~(\ref{ha2}) 
are also given in terms of the confluent Heun functions,
and corresponding analytical conditions for determining
the energy spectrum are obtained.
Similar to the parity-symmetric quantum Rabi model~\cite{Zhong},
the energy spectrum of the non-parity-symmetric quantum Rabi model
includes two parts, namely the regular and exceptional parts. 
The regular part can be be determined by the zeros of the Wronskian for the corresponding solutions. 
However, the exceptional part has to be derived from termination conditions for the confluent 
Heun functions in the corresponding solutions.

In particular, we show in Sec.~3.2 how the energy spectrum changes with the parity-symmetry-breaking term. 
At $\epsilon=0$, a two-fold energy degeneracy corresponds to the exceptional points. 
As $\epsilon$ is increased from $0$ to $\omega/2$, the pairs of initially degenerate eigenvalues for the exceptional points 
at $\epsilon=0$ gradually separate. At $\epsilon=\omega/2$, the separated exceptional points collide and form new degenerate points of energy.

\subsection{Eigenvalue problem}

Following our recent method~\cite{Zhong}, the eigenstate $\left|\psi\right\rangle$ 
of the generalized quantum Rabi model $H_R^\epsilon$ can be expressed as
\begin{equation}
\left|\psi\right\rangle =\psi_1 (a^{\dagger})\left|0\right\rangle \left|\downarrow\right\rangle 
+\psi_2(a^{\dagger})\left|0\right\rangle \left|\uparrow\right\rangle,
\label{state}
\end{equation}
where $\psi_{1,2}$ are analytical functions of the creation operator $a^{\dagger}$, 
and $\left|0\right\rangle$ is the vacuum state. 
Substituting this expansion into the Schr\"{o}dinger equation $H|\psi\rangle=E|\psi\rangle$, we get the eigenvalue equation
\begin{equation}
([H,\psi_1]+\psi_1H-E\psi_1) \left|0\right\rangle \left|\downarrow\right\rangle  
+([H,\psi_2]+\psi_2H-E\psi_2) \left|0\right\rangle \left|\uparrow\right\rangle =0.
\end{equation}
By applying the well-known relations $a\left|0\right\rangle =0$, $[a^{\dagger},\psi_{1,2}]=0$ 
and $[a,\psi_{1,2}]=d\psi_{1,2}/da^{\dagger}$, one finds that the two components $\psi_{1,2}$ 
obey coupled operator-type differential equations of the form
\begin{eqnarray}
\omega z\frac{{d\psi_1}}{{d z}}+g\left(z+\frac{{d}}{{d z}}\right)\psi_2+\Delta \psi_1+\epsilon\psi_2 &=&E\psi_1,\\
\omega z\frac{{d\psi_2}}{{d z}}+g\left(z+\frac{{d}}{{d z}}\right)\psi_1-\Delta \psi_2+\epsilon\psi_1 &=&E\psi_2.
\end{eqnarray}
Obviously, all terms in this equation are mutually commutable. 
Similar to the analytical treatment for Bose-Hubbard Hamiltonians~\cite{Wu1, Wu2}, 
the above operator-type differential equations can be regarded as c-number differential equations.

Introducing the linear combinations $\psi_{\pm}=\frac{1}{2} \left(\psi_1\pm\psi_2\right)$, we have another pair of coupled equations for $\psi_+(z)$ and $\psi_-(z)$,
\begin{eqnarray}
\frac{{d\psi_+}}{{dz}}  &=& \frac{E-\epsilon-g z}{z+g}\psi_+- \frac{\Delta}{z+g}\psi_-,\label{ceqa}\\
\frac{{d\psi_-}}{{dz}}  &=& \frac{E+\epsilon+g z}{z-g}\psi_--
\frac{\Delta}{z-g}\psi_+.\label{ceqb}
\end{eqnarray}
Here and below, for brevity, we take $\omega=1$.

The above differential equations can be solved exactly following~\cite{Zhong} (the details are given in Appendix A). 
We obtain two sets of analytical solutions for $\psi_+$ and $\psi_-$,
\begin{eqnarray}
\psi_+^1(z)&=&e^{-g z} \, \textrm{HC} \left(\alpha_1,\beta_1,\gamma_1,\delta_1,\eta_1, \frac{g-z}{2g}\right), \label{af1a}\\
\psi_-^1(z)&=&\frac{\Delta e^{-g z}} {E+g^2+\epsilon} \textrm{HC} \left(\alpha_2,\beta_2,\gamma_2,\delta_2,\eta_2, \frac{g-z}{2g}\right), \label{af2a}
\end{eqnarray}
and
\begin{eqnarray}
\psi_+^2(z)&=&\frac{\Delta e^{g z}} {E+g^2-\epsilon} \textrm{HC} \left(\alpha_1,\gamma_1,\beta_1,-\delta_1,\eta_1+\delta_1, \frac{g+z}{2g}\right),\label{af1b}\\
\psi_-^2(z)&=&e^{g z} \, \textrm{HC} \left(\alpha_2,\gamma_2,\beta_2,-\delta_2,\eta_2+\delta_2, \frac{g+z}{2g} \right),
\label{af2b}
\end{eqnarray}
where 
\begin{equation}
\textrm{HC}(\alpha,\beta,\gamma,\delta,\eta,x) =\sum_{n=0}^{\infty} h_n x^n \label{Heun}
\end{equation}
 is the confluent Heun function~\cite{Ronveaux, Slavyanov}. 
 The coefficients $h_n$ of this special function are determined by a three-term recurrence relation (see equation~(\ref{recurrence})). 
 In comparison with the first set of analytical solutions $(\psi_+^1(z), \psi_-^1(z))$, 
 in the second set of analytical solutions $(\psi_+^2(z), \psi_-^2(z))$, 
 $\beta_{j}$ is interchanged with $\gamma_{j}$, $\delta_{j}$ changes sign, and $\eta_{j}$ becomes $\eta_{j}+\delta_{j}$, where $j=1$ and $2$.

\subsection{Energy spectrum}

We now determine the analytical energy spectrum for the two sets of analytical solutions (\ref{af1a})-(\ref{af2b}).
 We have shown that although the sets
 of analytical solutions $(\psi_+^1, \psi_-^1)$ and
 $(\psi_+^2, \psi_{-}^{2})$ seem to have different forms,
 they are actually two linearly dependent solutions (details are given in Appendix B).
Since the two solutions are  linearly dependent, their Wronskian must be zero \cite{Slavyanov}. 
We therefore have (see equation (\ref{wron}) in  Appendix B)
\begin{equation}
W_+^\epsilon(E,z) := \psi_{+}^{2}\frac{d\psi_{+}^{1}}{dz} -\psi_{+}^{1}\frac{d\psi_{+}^{2}}{dz}=0,
\end{equation}
and
\begin{equation}
W_-^\epsilon(E,z) := \psi_{-}^{2}\frac{d\psi_{-}^{1}}{dz} -\psi_{-}^{1}\frac{d\psi_{-}^{2}}{dz}=0.
\end{equation}

According to~\cite{Braak, Zhong}, the conditions $W_{\pm}^\epsilon(E,z)=0$ hold
for arbitrary values of $z$ if and only if $E$
corresponds to an energy eigenvalue of the generalized quantum Rabi model.
Here we choose $z=0$ as done in~\cite{Braak, Zhong}.
The zeroes of $W_{\pm}^\epsilon(E,z)$ give the regular parts of the energy spectrum. 
Note that because $\psi_+^1(-z, -\epsilon)=\psi_-^2(z,\epsilon)$ and $\psi_-^1(-z, -\epsilon)=\psi_+^2(z,\epsilon)$, 
we have $W_{+}^{\epsilon}(E,z) =W_{-}^{-\epsilon}(E,-z)$, so we only need discuss $W_{+}^{\epsilon}(E,z)$.

\begin{figure}[htb]
\begin{center}
\includegraphics[width=0.8\columnwidth]{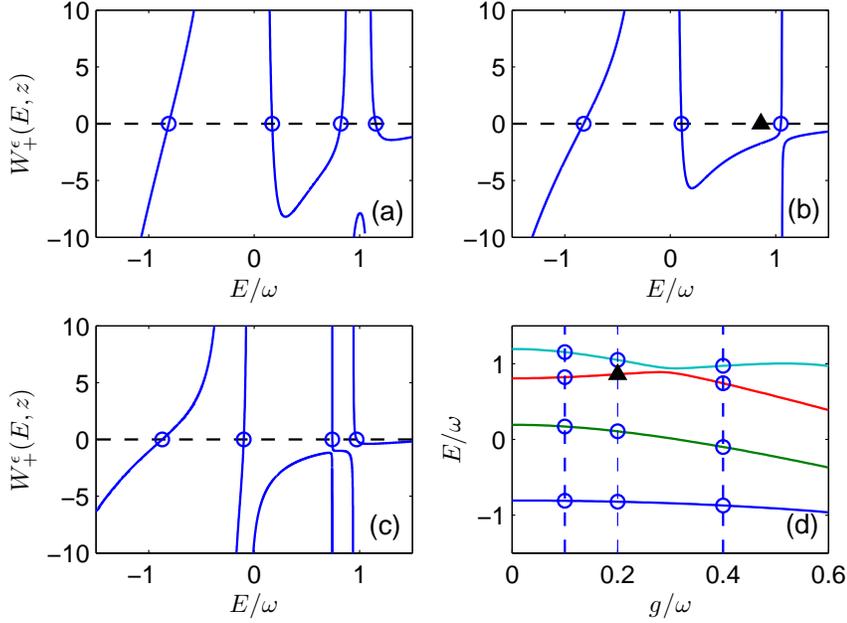}
\caption{Plots of the Wronskian $W_{+}^\epsilon(E,z)$ as a function
of $E/\omega$ for $\omega=1$, $\Delta/\omega=0.8$, $\epsilon/\omega=0.1$ 
and three different values of $g/\omega$: (a) 0.1, (b) 0.2 and (c) 0.4. 
Part (d) shows the corresponding energy spectrum as a function of
$g/\omega$. From left to right, the vertical dashed lines in (d) correspond
to $g/\omega=0.1$, $0.2$ and $0.4$, respectively. The circles
represent the zeros of $W_{+}^\epsilon(E,z)$. The black triangle
denotes the exceptional energy eigenvalue which does not correspond
to $W_{+}^\epsilon(E,z)=0$.}\label{fig1}
\end{center}
\end{figure}

The Wronskian $W_{+}^{\epsilon}(E,z)$ is shown as a function of $E/\omega$ for typical values of the parameters in  Fig.~\ref{fig1}. 
For simplicity, we display only the finite range $-1.5\leq E/\omega\leq 1.5$. 
It is observed that $W_{+}^{\epsilon}(E,z)$ has zeros as a function of $E/\omega$, marked by circles in the figure.
In Fig.~\ref{fig1}(d), we show the four lowest energy levels obtained by numerical diagonalization. 
The circles in Fig.~\ref{fig1}(d) correspond to the regular energy eigenvalues shown in Fig.~\ref{fig1}(a)-(c). 
It is clearly seen that for $g/\omega=0.1$ and $0.4$, the zeros of $W_{\pm}^{\epsilon}(E,z)$ give all four
lowest energy levels.

However, for $g/\omega=0.2$, the third -- exceptional -- energy level in Fig.~\ref{fig1}(d), marked by a black triangle, 
does not correspond to a zero point of $W_{+}^{\epsilon}(E,z)$ in Fig.~\ref{fig1}(b).
This indicates that there exist exceptional energy eigenvalues which cannot be determined by the zeros of
$W_{\pm}^\epsilon(E,z)$. 
Actually, the exceptional energy eigenvalues correspond to the analytical solutions with one
divergent confluent Heun function and  one finite series, as we shall further explain below.

As shown in Appendix C, there are a set of conditions
\begin{eqnarray}
&&h_{N+1}=0, \label{conditionb}\\
&&\delta=-(N+(\gamma+\beta+2)/2)\alpha,\label{conditiona}
\end{eqnarray}
with an integer $N\geq 0$. Under these conditions, the confluent
Heun function may be terminated as a polynomial with $N$ terms, as in equation (\ref{finite}). 
By applying the conditions (\ref{conditionb})
and (\ref{conditiona}), one can give the explicit expressions for the
corresponding energy eigenvalues.

For the first set of analytical solutions $(\psi_{+}^{1}, \psi_{-}^{1})$, 
according to equations (\ref{conditionb}) and (\ref{conditiona}), we have
${E_{+}}/{\omega}=N_{+}-{g^2}/{\omega^2}+{\epsilon}/{\omega}$ and $h_{N_{+}+1}(g,\Delta,\epsilon)=0$ for
$\psi_{+}^{1}(z)$, and
${E_{-}}/{\omega}=N_{-}-{g^2}/{\omega^2}+1
+{\epsilon}/{\omega}$ and $h_{N_{-}+1}(g,\Delta,\epsilon)=0$ for $\psi_{-}^{1}(z)$. 
As $\psi_{+}^{1}(z)$ and $\psi_{-}^{1}(z)$ are a
set of analytical solutions for the same eigenstate, we have
$E_{+}=E_{-}=E$ which requires $N_{+}=N_{-}+1$. 
For example, for the simplest non-trivial case of $N_{+}=1$ and $N_{-}=0$, we have the
energy eigenvalue
\begin{equation}
\frac{E}{\omega}=1-\frac{g^2}{\omega^2} +\frac{\epsilon}{\omega},\label{exam1}
\end{equation}
for the parameter relation
\begin{equation}
\frac{\Delta^2}{\omega^2}+\frac{4g^2}{\omega^2}
=1+\frac{2\epsilon}{\omega}. \label{exam1con}
\end{equation}
The analytical solutions for this specific case are
\begin{eqnarray}
\psi_+^{1}(z)&=&\left(1-\frac{2g^2}{1+2\epsilon} +\frac{2gz}{1+2\epsilon}\right) e^{-g z},\\
\psi_-^{1}(z)&=&\frac{\Delta}{1+2\epsilon}e^{-g z}.
\end{eqnarray}

Using the relations $\psi_{1}=\psi_{+}+\psi_{-}$ and $\psi_{2}=\psi_{+}+\psi_{-}$ we obtain the explicit form for  the corresponding eigenstate,
\begin{eqnarray}
\left|\psi\right\rangle&=& e^{g^2/2} \, \left[ \left(1+\frac{\Delta-2g^2}{1+2\epsilon}\right) \left|-g\right\rangle 
+\sqrt{L_1(-g^2)} \frac{2g}{1+2\epsilon} \left|-g,1\right\rangle\right] \left|\downarrow\right\rangle \nonumber\\
&&+ e^{g^2/2} \,
\left[ \left(1-\frac{\Delta+2g^2}{1+2\epsilon} \right) \left|-g\right\rangle +\sqrt{L_1(-g^2)} \frac{2g}{1+2\epsilon} \left|-g,1\right\rangle\right] \left|\uparrow\right\rangle,
\end{eqnarray}
for which $\left|-g\right\rangle$ denotes the coherent state, 
$\left|-g,m\right\rangle$ denotes the photon added coherent state, 
and $L_m(x)$ is the Laguerre polynomial of order $m$. 
If  $N_{+}=2$ and $N_{-}=1$, we have the energy eigenvalue
\begin{equation}
\frac{E}{\omega}=2-\frac{g^2}{\omega^2} +\frac{\epsilon}{\omega},\label{exam2}
\end{equation}
for the parameter relation
\begin{equation}
\frac{64g^2}{\omega^2} +\frac{\Delta^4}{\omega^4}
+\frac{4\Delta^2}{\omega^2}+4
-\left(\frac{16g^2}{\omega^2}+\frac{3\Delta^2}{\omega^2}
-\frac{8\epsilon}{\omega}-6\right)^2=0.\label{exam2con}
\end{equation}
However, the forms of the corresponding analytical solutions are more complicated, and thus we do not give them here.

Similarly, for the second set of analytical solutions $(\psi_{+}^{2}, \psi_{-}^{2})$, 
we have ${E_{+}}/{\omega}=N_{+}+1-{g^2}/{\omega^2} -{\epsilon}/{\omega}$ and 
$h_{N_{+}+1}(g,\Delta,\epsilon)=0$ for $\psi_+^2$, and ${E_{-}}/{\omega}=N_{-}-{g^2}/{\omega^2} -{\epsilon}/{\omega}$ 
and $h_{N_{-}+1}(g,\Delta,\epsilon)=0$ for $\psi_-^2$. 
In this situation, the same energy eigenvalue $E_{+}=E_{-}=E$ for $\psi_+^2$ and $\psi_-^2$ requires $N_{-}=N_{+}+1$. 
For the simplest case of $N_{+}=0$ and $N_{-}=1$, we have the energy eigenvalue
\begin{equation}
\frac{E}{\omega}=1- \frac{g^2}{\omega^2} -\frac{\epsilon}{\omega}, \label{exam3}
\end{equation}
for the parameter relation
\begin{equation}
\frac{\Delta^2}{\omega^2}+\frac{4g^2}{\omega^2}
=1-\frac{2\epsilon}{\omega}.\label{exam3con}
\end{equation}
The corresponding analytical solutions are
\begin{eqnarray}
\psi_+^{2}(z) &=& \frac{\Delta}{1-2\epsilon}e^{g z},\\
\psi_-^{2}(z) &=& \left(1-\frac{2g^2}{1-2\epsilon} -\frac{2gz}{1-2\epsilon}\right) e^{g z},
\end{eqnarray}
with corresponding eigenstate 
\begin{eqnarray}
\left|\psi\right\rangle&=& e^{g^2/2} \, \left[ \left(1+\frac{\Delta-2g^2}{1-2\epsilon}\right) 
\left|g\right\rangle -\sqrt{L_1(-g^2)}\frac{2g}{1-2\epsilon} \left|g,1\right\rangle\right] \left|\downarrow\right\rangle  \nonumber\\
&&- e^{g^2/2} \, \left[ \left(1-\frac{\Delta+2g^2}{1-2\epsilon} \right) 
\left|g\right\rangle -\sqrt{L_1(-g^2)}\frac{2g}{1-2\epsilon} \left|g,1\right\rangle\right] \left|\uparrow\right\rangle.
\end{eqnarray}
In the case of $N_+=1$ and $N_-=2$ the energy eigenvalue is 
\begin{equation}
\frac{E}{\omega}=2-\frac{g^2}{\omega^2}-\frac{\epsilon}{\omega}, \label{exam4}
\end{equation}
for the parameter relation
\begin{equation}
\frac{64g^2}{\omega^2} +\frac{\Delta^4}{\omega^4}
+\frac{4\Delta^2}{\omega^2}+4
-\left(\frac{16g^2}{\omega^2}+\frac{3\Delta^2}{\omega^2}
+\frac{8\epsilon}{\omega}-6\right)^2=0.\label{exam4con}
\end{equation}

It is evident that for the eigenstates of the terminated confluent Heun functions, 
from equations (\ref{exam1}), (\ref{exam2}), (\ref{exam3}) and (\ref{exam4}), 
their energy eigenvalues can be expressed in the unified form
\begin{equation}
\frac{E^{\pm}_{N}}{\omega} = N - \frac{g^2}{\omega^2} \pm \frac{\epsilon}{\omega}, \label{EU}
\end{equation}
with positive integers $N$. 
It is well-known that the parity-symmetric quantum Rabi model ($\epsilon=0$) 
exhibits Judd isolated exact analytical solutions under specific relations between 
${\Delta}/{\omega}$ and ${g}/{\omega}$~\cite{Judd}. 
Here our analytical results show that, for the non-parity-symmetric quantum Rabi model 
($\epsilon \neq 0$), there still exist similar isolated exact solutions.
These analytical solutions are holomorphic over the whole complex plane~\cite{Swain,kus,Reika,Reikb,Koc}. 
Recall that in Fig.~\ref{fig1}~(b) and (d), we saw an exceptional point which does not correspond to a 
zero of $W_{+}^{\epsilon}(E,z)$. 
The particular chosen set of parameters 
($\omega=1$, ${g}/{\omega}=0.2$, ${\Delta}/{\omega}=0.8$, ${\epsilon}/{\omega}=0.1$) satisfy the  relation
${\Delta^2}/{\omega^2}+{4g^2}/{\omega^2}=1-{2\epsilon}/{\omega}$. 
Thus, according to equation~(\ref{exam3}), this exact energy eigenvalue is 
${E}/{\omega}=1-{g^2}/{\omega^2}-{\epsilon}/{\omega}=0.86$, which is the exceptional value marked by the black triangle in Fig.~\ref{fig1}.

In general the energy separation between exceptional points with the same integer
$N$ is determined by the parity-breaking term. 
According to equation~(\ref{EU}), for two exceptional points with the same $N$,
the energy separation is given by 
\begin{equation}
E^{+}_{N} - E^{-}_{N} = 2 \epsilon. 
\end{equation}
This means that the two exceptional points gradually separate as the parity-breaking term $\epsilon$  increases.

For some specific values of $\epsilon$, exceptional points with different $N$ may form degenerate points with the same energy. Given
\begin{equation}
\frac{E^{+}_{N_1}}{\omega}=N_1-\frac{g^2}{\omega^2} +\frac{\epsilon}{\omega},
\end{equation}
and
\begin{equation}
\frac{E^{-}_{N_2}}{\omega}=N_{2}-\frac{g^2}{\omega^2} -\frac{\epsilon}{\omega},
\end{equation}
i.e., two exceptional points with opposite sign of $\epsilon$,
it is easy to find $E^{+}_{N_1}=E^{-}_{N_2}$ when $\epsilon=\frac{1}{2}(N_2-N_1)\omega$.
The two exceptional points thus form a two-fold degenerate point of energy when $\epsilon$ is an integer multiple of $\frac{1}{2}\omega$. 
These analytical results give a direct demonstration of Braak's argument for the energy level crossings 
at ${\epsilon}/{\omega}={N}/{2}$~\cite{Braak}.

To clearly illustrate our results for the two-fold degenerate
exceptional points, consider as an example the case $N_1=1$ and $N_2=2$. 
From equations~(\ref{exam1}) and (\ref{exam4}), 
$E^{+}_{1}=E^{-}_{2} =\left(\frac{3}{2} -\frac{g^2}{\omega^2}\right)\omega$ if
${\epsilon}/{\omega}=\frac{1}{2}$. 
More importantly, the
two-fold degeneracy requires that the parameter relations for
$E^{+}_{1}=\left(\frac{3}{2} -\frac{g^2}{\omega^2}\right)\omega$ and
$E^{-}_{2}=\left(\frac{3}{2} -\frac{g^2}{\omega^2}\right)\omega$ are
valid simultaneously. 
From equation~(\ref{exam1con}) the parameter relation for $E^{+}_{1}$ is 
\begin{equation}
\frac{\epsilon}{\omega} =\frac{1}{2} \left(\frac{\Delta^2}{\omega^2}+\frac{4g^2}{\omega^2}-1\right). 
\label{rela}
\end{equation}
Substituting this equation in the parameter relation (\ref{exam4con}) for $E^{-}_{2}$ yields
\begin{eqnarray}
\frac{64g^2}{\omega^2} +\frac{\Delta^4}{\omega^4} +\frac{4\Delta^2}{\omega^2}+4 -\left(\frac{16g^2}{\omega^2} +\frac{3\Delta^2}{\omega^2} +\frac{8\epsilon}{\omega} -6\right)^2 \nonumber\\
=-16\left(\frac{\Delta^2}{\omega^2} +\frac{4g^2}{\omega^2}-2\right) \left(\frac{3\Delta^2}{\omega^2} +\frac{16g^2}{\omega^2}-3\right) \label{degcon1}
\end{eqnarray} 
which clearly vanishes when ${\epsilon}/{\omega}=\frac{1}{2}$, for which the parameter relation (\ref{rela}) for $E^{+}_{1}$ becomes 
\begin{equation}
\frac{\Delta^2}{\omega^2} +\frac{4g^2}{\omega^2}=2. \label{degcon}
\end{equation}
This ensures the validity of the parameter relation (\ref{exam4con}) for $E^{-}_{2}$. 
We have thus seen that under the conditions ${\epsilon}/{\omega}=\frac{1}{2}$ and ${\Delta^2}/{\omega^2}+{4g^2}/{\omega^2}=2$, 
the simultaneous validity of the parameter relations for $E^{+}_{1}$ and $E^{-}_{2}$ indicates the degeneracy of the two exceptional points. 
In principle, we can obtain all possible conditions for level-crossings with the help of the termination conditions for the confluent Heun functions.

\begin{figure}[tb]
\begin{center}
\includegraphics[width=0.8\columnwidth]{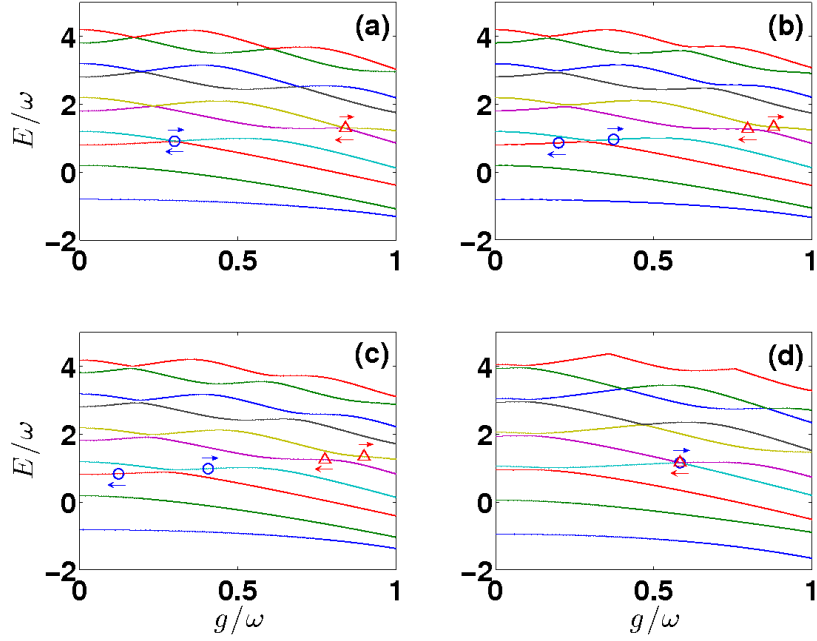}
\caption{Energy spectrum of the non-parity-symmetric quantum Rabi model as a function of ${g}/{\omega}$ with $\omega=1$ and ${\Delta}/{\omega}=0.8$ 
for the values (a) ${\epsilon}/{\omega} =0$ (b) ${\epsilon}/{\omega} = 0.1$ (c) ${\epsilon}/{\omega} = 0.15$ and (d) ${\epsilon}/{\omega} = 0.5$. 
The circles and triangles denote the energy eigenvalues given by equations (\ref{exam1}) and (\ref{exam4}), respectively.}
\label{fig2}
\end{center}
\end{figure}

In Fig.~\ref{fig2} we show the energy spectrum of the ten lowest energy levels as a function of ${g}/{\omega}$ 
for different values of ${\epsilon}/{\omega}$.
For the parity-symmetric case (${\epsilon}/{\omega}=0$) the energy spectrum shows several level crossings
at the exceptional points. 
These level crossings are related to Judd's isolated solutions. 
Here we discuss two sets of them, marked by circles and triangles. 
The circles denote the exact energy eigenvalues given by equations (\ref{exam1}) and  (\ref{exam3}), 
and the triangles denote the exact energy eigenvalues given by equations (\ref{exam2}) and (\ref{exam4}). 
For ${\epsilon}/{\omega}=0$, the energy eigenvalues are two-fold degenerate, as shown in Fig.~\ref{fig2}(a). 
As ${\epsilon}/{\omega}$ increases, the two-fold degeneracy is broken and the energy level 
crossings become anti-crossings, see Fig.~\ref{fig2}(b) and Fig.~\ref{fig2}(c). 
The arrows denote the direction of separation of the exceptional points as ${\epsilon}/{\omega}$ increases. 
In particular, when ${\epsilon}/{\omega} \to \frac{1}{2}$, the two lowest exceptional points meet at 
${g}/{\omega} =\frac{1}{2}\sqrt{2-\frac{\Delta^2}{\omega^2}} \approx 0.583$ (see equation (\ref{degcon})) 
and form a new two-fold degenerate energy point, as indicated in Fig.~\ref{fig2}(d).

We now explain the occurrence of the exceptional points by analyzing the
confluent Heun functions in $(\psi_+^{1}, \psi_-^{1})$ and $(\psi_+^{2}, \psi_-^{2})$. 
According to equations (\ref{recurrence}) and (\ref{a1}), 
the confluent Heun function $\textrm{HC}(\alpha,\beta,\gamma,\delta,\eta,x)$ 
becomes divergent if $A(N)=0$, which requires a negative integer $\beta=-N$. 
On the other hand, we know that under the conditions (\ref{conditionb}) and (\ref{conditiona}), 
the confluent Heun function $\textrm{HC}(\alpha,\beta,\gamma, \delta,\eta, \frac{g-z}{2g})$ is a finite series.
Now in comparing the two confluent Heun functions in $\psi_+^1(z)$ and $\psi_+^2(z)$, $\gamma_1$ is interchanged  with $\beta_1$. 
According to the truncation conditions (\ref{conditionb}) and (\ref{conditiona}) and the parameter relations given in the text below equation (\ref{heun1}), 
if $\psi_+^1(z)$ has a truncated confluent Heun function $\textrm{HC}(\alpha_1,\beta_1,\gamma_1, \delta_1,\eta_1, \frac{g-z}{2g})$, 
we have ${E}/{\omega}=N_1-{g^2}/{\omega^2} +{\epsilon}/{\omega}$, 
$\beta_1=-({E}/{\omega}+{g^2}/{\omega^2} +{\epsilon}/{\omega}+1) =-(N_1 +1+2 {\epsilon}/{\omega})$ 
and $\gamma_1=-({E}/{\omega} -{\epsilon}/{\omega} +{g^2}/{\omega^2})=-N_1$. 
Therefore, since $\gamma_1=-N_1$ is a negative integer, the confluent Heun function 
$\textrm{HC}(\alpha_1,\gamma_1,\beta_1, -\delta_1, \eta_1+\delta_1, \frac{g+z}{2g})$ in $\psi_+^2(z)$ 
is divergent and $\psi_+^2(z)$ becomes a non-physical solution. 
This means that, corresponding to the energy eigenvalue 
${E}/{\omega}=N_1-{g^2}/{\omega^2} +{\epsilon}/{\omega}$, $\psi_+^1(z)$ 
has a truncated confluent Heun function and $\psi_+^2(z)$ is a non-physical solution.

Similarly, according to the truncation conditions (\ref{conditionb}) and (\ref{conditiona}) 
and the parameter relations given in the text below (\ref{heun1}), 
if $\psi_+^2(z)$ has a truncated confluent Heun function 
$\textrm{HC}(\alpha_1,\gamma_1,\beta_1, -\delta_1, \eta_1+\delta_1, \frac{g+z}{2g})$, 
we have ${E}/{\omega}=N_2-{g^2}/{\omega^2} -{\epsilon}/{\omega}$, $\beta_1=-N_2$ 
and $\gamma_1=-(N_2-2{\epsilon}/{\omega})$. 
Therefore, since $\beta_1=-N_2$ is a negative integer, the confluent Heun function 
$\textrm{HC}(\alpha_1,\beta_1,\gamma_1, \delta_1,\eta_1,\frac{g-z}{2g})$ in $\psi_+^1(z)$ 
is divergent and $\psi_+^1(z)$ becomes a non-physical solution. 
This means that, corresponding to the energy eigenvalue 
${E}/{\omega}=N_2-{g^2}/{\omega^2}-{\epsilon}/{\omega}$, $\psi_+^2(z)$ 
has a truncated confluent Heun function and $\psi_+^1(z)$ is a non-physical solution.

Summarizing the above analysis of the confluent Heun functions, 
for the exceptional points whose energy eigenvalues are expressed as 
${E}/{\omega}=N-{g^2}/{\omega^2} + {\epsilon}/{\omega}$ or 
${E}/{\omega}=N-{g^2}/{\omega^2} - {\epsilon}/{\omega}$, 
one of the two confluent Heun functions for a specific $N$ is a finite series and the other is divergent. 
Therefore, in this case the two sets of analytical solutions $(\psi_{+}^{1},\psi_{-}^{1})$ and $(\psi_{+}^{2},\psi_{-}^{2})$ 
are not linear dependent solutions and the corresponding energy eigenvalues do not correspond to the zeros of 
the Wronskian $W_{\pm}^\epsilon(E,z)$ (recall Fig.~\ref{fig1}).

\section{Conclusion}

In conclusion, we have presented an analytical energy spectrum for hybrid mechanical systems described by non-parity-symmetric quantum Rabi models. 
We have given their analytical solutions in terms of the confluent Heun functions and derived the conditions for determining the energy spectrum. 
Due to the absence of the parity symmetry, the Hilbert space does not separate into two subspaces and there is no energy-level degeneracy, 
except for $\epsilon=N\omega/2$ with integer numbers $N$. 
Similar to the parity-symmetric quantum Rabi model, the energy spectrum still includes regular and exceptional parts. 
The regular parts are determined by the zeros of the Wronskian for the corresponding analytical solution. 
Under some specific conditions, the systems admit exactly analytical energy levels and eigenstates in closed form. 
They are systematically found from the conditions under which the confluent Heun functions are terminated as a polynomial. 
These analytical results correspond to the exceptional parts of the energy spectrum. 
The analytical conditions for the level crossings for the hybrid mechanical systems have also been found.

We note that there exist two other different methods for determining the energy spectrum of the generalized Rabi models. 
In Ref.~\cite{Braak}, the energy spectrum has been obtained by embedding this model into a larger system possessing a parity symmetry. 
In Ref.~\cite{Hu}, the Bogoliubov operator method has been used. 
The Bogoliubov operator method is  essentially similar to our method~\cite{Zhong}, but the condition for determining the energy spectrum 
is different from the one given here. 
Naturally, all methods give the same regular parts of the energy spectrum. 
With the help of the confluent Heun functions, it seems that our method can more simply find all isolated eigenstates 
and their energy eigenvalues associated with the exceptional parts of the energy spectrum in a systematic way. 
In particular, these isolated exact results allow us to better understand the conditions for the level crossings of this non-parity-symmetric model~\cite{Braak}.

Beyond the hybrid mechanical systems, our analytical results hold for all systems described by the generalized quantum Rabi model Hamiltonian (6). 
The model can not only be realized by the superconducting devices outlined in Sec.~2, 
but also can be realized by several other systems, such as a single ultracold ion in a spin-dependent Paul trap~\cite{LeeNJP}, 
individual ultracold atoms in spin-dependent optical lattices~\cite{Forster}, and a suspended carbon nanotube quantum dot~\cite{Palyi}. 
For example, for the case of a single trapped ultracold ion, the term $\omega a^{\dagger}a$ is created by a linear Paul trap, 
the coupling term $g\sigma_z(a^{\dagger}+a)$ can be realized by a gradient magnetic field, 
and the other two terms $-\Delta\sigma_x$ and $\epsilon\sigma_z$ 
may be realized by Raman coupling between the two involved hyperfine levels of the ion.
Our results should be of use to calculate relevant physical properties for such models.

\ack
This work is supported by the NBRPC under grant no 2012CB821305, 
the NNSFC under grant nos 11075223, 11374375, 11147021 and 11375059, 
the Ph.D. Programs Foundation of Ministry of Education of China under Grant No. 20120171110022, 
and the Hunan Provincial Natural Science Foundation under grant no 12JJ4010. 
MTB is supported by the 1000 Talents Program of China. His work is also supported by the Australian Research Council through grant DP130102839. 
He also thanks All Souls College and the Rudolf Peierls Centre for Theoretical Physics, Oxford for kind hospitality and support.

\appendix

\section{Derivation of equations (\ref{af1a})-(\ref{af2b})}

Here we give the details of the derivation of equations (\ref{af1a})-(\ref{af2b}). 
By eliminating $\psi_+$ and $\psi_-$ from equations (\ref{ceqa}) and (\ref{ceqb}), 
we obtain two second-order differential equations for $\psi_+(z)$ and $\psi_-(z)$, 
\begin{eqnarray}
\frac{{d^2\psi_+}}{{dz^2}} +p_+(z)\frac{{d\psi_+}}{{dz}} +q_+(z)\psi_+=0,\label{seqa}\\
\frac{{d^2\psi_-}}{{dz^2}} +p_-(z)\frac{{d\psi_-}}{{dz}} +q_-(z)\psi_-=0,\label{seqb}
\end{eqnarray}
with
\begin{eqnarray}
p_{\pm}(z) &=& \frac{(1-2E-2g^2)z-(2\epsilon\pm1)g}{z^2-g^2},\nonumber\\
q_{\pm}(z) &=&\frac{-g^2 z^2-(2\epsilon\mp1)g z+E^2-g^2 -\epsilon^2-\Delta^2}{z^2-g^2}.\nonumber
\end{eqnarray}
Equations (\ref{seqa}) and (\ref{seqb}) admit two different types of solutions, which we write as
\begin{eqnarray}
&\textrm{Type-I:}~~~& \psi_{\pm}(z)=e^{-gz} \phi_{1,2}(x_{1}),\\
&\textrm{Type-II:}~~& \psi_{\pm}(z)=e^{gz} \phi_{3,4}(x_2),
\end{eqnarray}
where $x_1=(g-z)/2g$ and $x_2=(g+z)/2g$.

Substituting $\psi_+(z)=e^{-gz} \phi_1(x_1)$ into equation (\ref{seqa}), leads to a confluent Heun equation for $\phi_1(x_1)$
\begin{equation}
\frac{d^2\phi_1}{dx_1^2} +\left(\alpha_1+\frac{\beta_1+1}{x_1} +\frac{\gamma_1+1}{x_1-1}\right) 
\frac{d\phi_1}{dx_1} +\frac{\mu_1 x_1+\nu_1}{x_1(x_1-1)}\phi_1=0,
\label{heun1}
\end{equation}
where $\mu_1=\delta_1+\alpha_1(\beta_1+\gamma_1+2)/2$ and $\nu_1=\eta_1+\beta_1/2+(\gamma_1-\alpha_1)(\beta_1+1)/2$. 
The other parameters are $\alpha_1=4g^2$, $\beta_1=-(E+\epsilon+g^2+1)$, 
$\gamma_1=-(E-\epsilon+g^2)$, $\delta_1=-2(1-2\epsilon)g^2$, 
and $\eta_1=-3g^4/2+(1-2E-4\epsilon)g^2/2 +(E^2+E-\epsilon^2+\epsilon-2\Delta^2+1)/2$. 
Now if $-\beta_1=(E+\epsilon+g^2+1)$ is not zero or a negative integer, 
the confluent Heun equation has two linearly independent local Frobenius solutions around $x_1=0$ of the form~\cite{Ronveaux,Slavyanov}
\begin{eqnarray}
\phi_1^1(x_1)&=& \textrm{HC}(\alpha_1,\beta_1,\gamma_1,\delta_1,\eta_1,x_1),\\
\phi_1^2(x_1)&=& x_1^{-\beta_1} \textrm{HC}(\alpha_1,-\beta_1,\gamma_1,\delta_1,\eta_1,x_1),
\end{eqnarray}
where $\textrm{HC}(\alpha_1,\beta_1,\gamma_1, \delta_1, \eta_1, x_1) =\sum_{n=0}^{\infty}h_nx_1^n$ is the confluent Heun function. 
The coefficients $h_n$ obey the three-term recurrence relation
\begin{equation}
h_n=B(n)h_{n-1}/A(n) +C(n) h_{n-2}/A(n)  \label{recurrence}
\end{equation}
for $n\geq 1$, with the initial conditions $h_0=1$ and $h_{-1}=0$. 
Here $A(n)$, $B(n)$, and $C(n)$ are given by 
\begin{eqnarray}
A(n)&=&(n+\beta_1)n,\label{a1}\\
B(n)&=&n^2+(\beta_1+\gamma_1-\alpha_1-1)n +\eta_1-\beta_1/2\nonumber\\
&&+(\gamma_1-\alpha_1)(\beta_1-1)/2, \label{b1}\\
C(n)&=&\delta_1+\alpha_1(\beta_1+\gamma_1)/2 +\alpha_1(n-1)\label{c1}.
\end{eqnarray}

If both $\phi_1^1(x_1)$ and $\phi_1^2(x_1)$ are physically acceptable solutions, 
they must be entire functions over the whole complex plane. 
The solution $\phi_1^1$ satisfies such a requirement, at least formally, 
since $\phi_1^1$ is a series for $x_1=(g-a^{\dagger})/2g$. 
However, $\phi_1^2$ represents a physical solution only when $-\beta_1$ 
is a positive integer because of the presence of the term $x_1^{-\beta_1}$. 
The parameter $-\beta_1=E+g^2+\epsilon+1$ is clearly related to the physical parameters $E$, $g$ and $\epsilon$. 
Only under special conditions for $E$, $g$, and $\epsilon$, is $-\beta_1$ a positive integer, and thus represents a physical solution. 
Therefore, in general, we use only $\phi_1^1$ to construct the final solution for $\psi_+(z)$
\begin{equation}
\psi_+^1(z)=C_1e^{-gz}
\textrm{HC} \left(\alpha_1,\beta_1,\gamma_1,\delta_1,\eta_1, \frac{g-z}{2g}\right),
\label{f1a}
\end{equation}
where $C_1$ is a constant to be determined.

For the transformation $\psi_-(z)=e^{-g z}\phi_2(x_1)$, we find after a straightforward calculation  
that $\phi_2(x_1)$ also satisfies a confluent Heun equation
\begin{equation}
\frac{d^2\phi_2}{dx_1^2}+\left(\alpha_2+\frac{\beta_2+1}{x_1} +\frac{\gamma_2+1}{x_1-1}\right) 
\frac{d\phi_2}{dx_1} +\frac{\mu_2 x_1+\nu_2}{x_1(x_1-1)}\phi_2=0,
\end{equation}
where $\mu_2=\delta_2+\alpha_2(\beta_2+\gamma_2+2)/2$ and $\nu_2=\eta_2+\beta_2/2+(\gamma_2-\alpha_2)(\beta_2+1)/2$. 
In this case, different parameters appear in the confluent Heun equation. 
They are given by $\alpha_2=4g^2$, $\beta_2=-(E+\epsilon+g^2)$, $\gamma_2=-(E-\epsilon+g^2+1)$, 
$\delta_2=2(1+2\epsilon)g^2$, and $\eta_2=-3g^4/2-(3+2E+4\epsilon)g^2/2 +(E^2+E-\epsilon^2-\epsilon-2\Delta^2+1)/2$. 
Similarly, we have  the solution for $\psi_-(z)$ in the form
\begin{equation}
\psi_-^1(z)=C_2e^{-gz}
\textrm{HC} \left(\alpha_2,\beta_2,\gamma_2,\delta_2,\eta_2, \frac{g-z}{2g}\right),
\label{f1b}
\end{equation}
where $C_2$ is  a constant. 
The two constants $C_1$ and $C_2$ are determined by the requirement that they must satisfy either equation (\ref{ceqa}) or (\ref{ceqb}). 
Here we use equation (\ref{ceqb}) due to the presence of the term $z-g$ in the denominators. 
Substituting $\psi_+(z)$ and $\psi_-(z)$ into equation (\ref{ceqb}), it is easy to find $C_2/C_1=\Delta/(E+g^2+\epsilon)$.

On the other hand, if we introduce another variable $x_2=(g+z)/2g=1-x_1$ and the transformation 
$\psi_{\pm}=e^{gz}\phi_{3,4}(x_2)$, one may find different forms of the solutions. 
After substitution of $\psi_{\pm}(z)= e^{gz} \, \phi_{3,4}(x_2)$ into equations (\ref{seqa}) and (\ref{seqb}), 
we find that $\phi_{3,4}(x_2)$ obey the confluent Heun equations
\begin{equation}
\frac{d^2\phi_{3,4}}{dx_2^2}+\left(\alpha_{3,4}+\frac{\beta_{3,4}+1}{x_2}
+\frac{\gamma_{3,4}+1}{x_2-1}\right) 
\frac{d\phi_{3,4}}{dx_2} +\frac{\mu_{3,4}x_2+\nu_{3,4}}{x_2(x_2-1)}\phi_{3,4}=0,
\end{equation}
where
$\mu_{3,4}=\delta_{3,4} +\alpha_{3,4} (\beta_{3,4}+\gamma_{3,4}+2)/2$ 
and $\nu_{3,4}=\eta_{3,4}+\beta_{3,4}/2 +(\gamma_{3,4}-\alpha_{3,4}) (\beta_{3,4}+1)/2$. 
The parameter relations are given by $\alpha_{3,4}=\alpha_{1,2}$, $\beta_{3,4}=\gamma_{1,2}$, 
$\gamma_{3,4}=\beta_{1,2}$, $\delta_{3,4}=-\delta_{1,2}$ and $\eta_{3,4}=\eta_{1,2}+\delta_{1,2}$. 
Using similar working as for $\psi_+^1$ and $\psi_-^1$, we obtain the analytical solutions for $\psi_+(z)$ and $\psi_-(z)$ in a different form,
\begin{eqnarray}
\psi_+^2(z)&=&D_1e^{gz} \textrm{HC} \left(\alpha_1,\gamma_1,\beta_1,-\delta_1, \eta_1+\delta_1, \frac{g+z}{2g}\right), \label{f2a}\\
\psi_-^2(z)&=&D_2e^{gz}
\textrm{HC} \left(\alpha_2,\gamma_2,\beta_2,-\delta_2, \eta_2+\delta_2, \frac{g+z}{2g} \right),
\label{f2b}
\end{eqnarray}
where $D_1/D_2=\Delta/(E+g^2-\epsilon)$ can be determined by substituting $\psi_+(z)$ and $\psi_-(z)$ into equation (\ref{ceqa}).

The constants $C_1$ and $D_2$ are determined by the normalization of the wavefunctions. 
Here for simplicity, we choose $C_1=D_2=1$. 
Therefore, equations (\ref{f1a}) and (\ref{f1b}) correspond to equations (\ref{af1a}) and (\ref{af2a}), 
and equations (\ref{f2a}) and (\ref{f2b}) correspond to equations (\ref{af1b}) and (\ref{af2b}).

\section{Demonstration of linear dependence between two sets of analytical solutions}

The two sets of analytical solutions $(\psi_+^1, \psi_-^1)$ and $(\psi_+^2, \psi_{-}^{2})$ seem to have different forms. 
However, it can be shown that they are actually linearly dependent. 
Here we take $\psi_{+}^1$ and $\psi_+^2$ as an example to demonstrate this relation. 
If we further change the variable $x=1-x_1$ in equation (\ref{heun1}), it becomes 
\begin{equation}
\frac{d^2\phi_1}{dx^2} +\left(-\alpha_1+\frac{\beta_1+1}{x-1} +\frac{-\gamma_1+1}{x}\right) \frac{d\phi_1}{dx} +\frac{-\mu_1 x+\mu_1+\nu_1}{x(x-1)}\phi_1=0
\end{equation}
which is another version of the confluent Heun equation.
This implies that under the transformation $x=1-x_1$, the parameters transform according to the relations 
$\alpha_1\rightarrow-\alpha_1$, $\beta_1\rightarrow\gamma_1$, $\gamma_1\rightarrow\beta_1$, 
$\delta_1\rightarrow-\delta_1$, and $\eta_1\rightarrow\eta_1+\delta_1$. It is evident that the solution for $\psi_+$ may also be expressed as
\begin{equation}
\psi_+^1=C_1' \, e^{-gz} \, 
\textrm{HC}(-\alpha_1,\gamma_1,\beta_1,-\delta_1, \eta_1+\delta_1,1-x_1)
\end{equation}
where $C_1'$ is a constant. 
With $x_2=1-x_1=(g+z)/2g$ and making use of the result~\cite{Slavyanov} 
$\textrm{HC}(\alpha,\beta,\gamma,\delta,\eta,z) =  e^{-\alpha}  \, \textrm{HC}(-\alpha,\beta,\gamma,\delta,\eta,z)$, 
we have
\begin{equation}
\psi_+^1=C_1'  \, e^{2g^2 + gz} \, 
\textrm{HC}\left(-\alpha_1,\gamma_1,\beta_1,-\delta_1, \eta_1+\delta_1,\frac{g+z}{2g}\right). 
\end{equation}
The argument is that $\psi_+^1$ and $\psi_+^2$ are thus linearly dependent, because 
we only performed the change of variable. 
The linear dependence relation also applies for $\psi_-^1$ and $\psi_-^2$.

For a pair of functions $\varphi_1(z)$ and $\varphi_2(z)$, their Wronskian is defined as
\begin{equation}
W(\varphi_1,\varphi_2) : =\varphi_1\frac{d\varphi_2}{dz} -\varphi_2\frac{d\varphi_1}{dz}. \label{wron}
\end{equation}
If $\varphi_1(z)$ and $\varphi_2(z)$ are linearly dependent, their Wronskian vanishes identically~\cite{Slavyanov}. 
Therefore, as $(\psi_+^1, \psi_-^1)$ and $(\psi_+^2, \psi_{-}^{2})$ are linearly dependent, their Wronskian must be zero.

\section{Derivation of equations (\ref{conditionb}) and (\ref{conditiona})}

Here we present the details of the derivation of the conditions (\ref{conditionb}) and (\ref{conditiona}). 
Under the condition 
\begin{equation}
h_{n}=0 \quad\textrm{~if~}\quad n>N,
\end{equation}
the confluent Heun function (\ref{Heun}) clearly reduces to a polynomial of $N$ terms,
\begin{equation}
\textrm{HC}(\alpha,\beta,\gamma,\delta,\eta,x) =\sum_{n=0}^{N}h_nx^N.   \label{finite}
\end{equation}
According to the recurrence relation (\ref{recurrence}), we must have
\begin{equation}
h_{N+2}=B(N+2)h_{N+1}/A(N+2) +C(N+2) h_{N}/A(N+2). 
\end{equation}
So given condition (\ref{conditionb}), namely $h_{N+1}=0$, the condition $h_{N+2}=0$ requires
\begin{equation}
C(N+2)=0
\end{equation}
which gives the condition (\ref{conditiona}), namely 
$\delta=-(N+(\gamma+\beta+2)/2)\alpha$.
From repeated use of the recurrence relation (\ref{recurrence}) with $h_{N+1}=0$ and $h_{N+2}=0$, 
it is easy to find that $h_{n}=0$ if $n>N$. 
Therefore, under the conditions given by equations (\ref{conditionb}) and (\ref{conditiona}), 
the confluent Heun function becomes a finite series.


\section*{References}
\begin{thebibliography}{10}

\bibitem{Treutlein}Treutlein P, Genes C, Hammerer K, Poggio M and Rabl P 2012 arXiv:1210.4151

\bibitem{Knobel}Knobel R G and Cleland A N 2003 Nature \textbf{424} 291

\bibitem{Zippilli}Zippilli S, Morigi G and Bachtold A 2009 Phys. Rev. Lett. \textbf{102} 096804

\bibitem{Bennett}Bennett S D, Cockins L, Miyahara Y, Gr\"{u}tter P and Clerk A A 2010 Phys. Rev. Lett. \textbf{104} 017203

\bibitem{Rugar}Rugar D, Budakian R, Mamin H J and Chui B W 2004 Nature \textbf{430} 329

\bibitem{Xue}Xue F,  Zhong L,  Li Y and Sun C P 2007 Phys. Rev. B \textbf{75} 033407


\bibitem{Armour} Armour A D, Blencowe M P, Schwab K C 2002 Phys. Rev. Lett. \textbf{88} 148301

\bibitem{Irish}Irish E K and Schwab K 2003 Phys. Rev. B \textbf{68} 155311

\bibitem{Martin}Martin I, Shnirman A, Tian L and Zoller P 2004 Phys. Rev. B \textbf{69} 125339

\bibitem{Xueb}Xue F, Wang Y D, Sun C P, Okamoto H, Yamaguchi H and Semba K 2007 New J. Phys. \textbf{9} 35

\bibitem{Wang}Wang Y J, Eardley M, Knappe S, Moreland J, Hollberg L and Kitching J 2006 Phys. Rev. Lett. \textbf{97} 227602

\bibitem{Hunger}Hunger D, Camerer S, H\"{a}nsch T W, K\"{o}nig D, Kotthaus J P, Reichel J and Treutlein P 2010 Phys. Rev. Lett. \textbf{104} 143002

\bibitem{Camerer}Camerer S, Korppi M, J\"{o}ckel M, Hunger D, H\"{a}nsch T W and Treutlein P 2011 Phys. Rev. Lett. \textbf{107} 223001

\bibitem{LeeNJP}Hu Y M, Yang W L, Xu Y Y, Zhou F, Chen L, Gao K L, Feng M and Lee C 2011 New J. Phys. \textbf{13} 053037

\bibitem{LeeSLZT}Zhang X, Huang J, Zhang Y, Gao K L and Lee C 2012 arXiv:1207.5585

\bibitem{Rabi}Rabi I I 1936 Phys. Rev. \textbf{49} 324
\nonum Rabi I I 1937 Phys. Rev. \textbf{51} 652

\bibitem{Braak}Braak D 2011 Phys. Rev. Lett. \textbf{107} 100401

\bibitem{Braakb}Braak D 2013 J. Phys. A \textbf{46} 175301
\nonum Braak D 2013 Ann. Phys. (Berlin) \textbf{525} L23

\bibitem{Wolf}Wolf F A, Kollar M and Braak D 2012 Phys. Rev. A \textbf{85} 053817

\bibitem{Wolfb}Wolf F A, Vallone F, Romero G, Kollar M, Solano E and Braak D 2013 Phys. Rev. A \textbf{87} 023835

\bibitem{Hirokawa}Hirokawa M and Hiroshima F 2012 arXiv:1207.4020

\bibitem{Chen}Chen Q H, Wang C, He S, Liu T and Wang K L 2012 Phys. Rev. A \textbf{86}  023822

\bibitem{Yu}Yu L, Zhu S, Liang Q, Chen G and Jia S 2012 Phys. Rev. A \textbf{86} 015803

\bibitem{Ziegler}Ziegler K 2012 J. Phys. A \textbf{45} 452001

\bibitem{Moroz}Moroz A 2012 EPL \textbf{100} 60010 
\nonum  Moroz A 2013 Ann. Phys. (N.Y.) \textbf{338} 319

\bibitem{Maciejewski}Maciejewski A J, Przybylska M and Stachowiak T 2012 arXiv:1210.1130

\bibitem{Zhong}Zhong H, Xie Q, Batchelor M T and Lee C 2013 J. Phys. A \textbf{46} 415302

\bibitem{Travenec}Trav\u{e}nec I 2012 Phys. Rev. A \textbf{85} 043805

\bibitem{Zhang}Zhang Y Z 2013 J. Math. Phys. \textbf{54} 102104 

\bibitem{Chilingaryan}Chilingaryan S A and Rodr\'{\i}guez-Lara B M 2013 J. Phys. A \textbf{46} 335301

\bibitem{Albert2012}Albert V V 2012 Phys. Rev. Lett. \textbf{108} 180401

\bibitem{Shen}Shen L T, Yang Z B, Lu M, Chen R X and Wu H Z 2013 arXiv:1306.2122

\bibitem{Tomka}Tomka M,  El Araby O, Pletyukhov M and Gritsev V 2013 arXiv:1307.7876

\bibitem{charge-qubit}Makhlin Y, Sch\"{o}n G and Shnirman A 2001 Rev. Mod. Phys. \textbf{73} 357

\bibitem{Wu1}Wu Y, Yang X and Xiao Y 2001 Phys. Rev. Lett. \textbf{86} 2200

\bibitem{Wu2}Wu Y and Yang X 2003 Phys. Rev. A \textbf{68} 013608

\bibitem{Judd}Judd B R 1979 J. Phys. C \textbf{12} 1685

\bibitem{Swain}Swain S 1973 J. Phys. A \textbf{6} 192

\bibitem{kus}Kus M and Lewenstein M 1986 J. Phys. A \textbf{19} 305

\bibitem{Reika}Reik H G and Doucha M 1986 Phys. Rev. Lett. \textbf{57} 787

\bibitem{Reikb}Reik H G, Lais P, St\"{u}tzle M E and Doucha M 1987 J. Phys. A \textbf{20} 6327

\bibitem{Koc}Ko\c{c} R, Koca M and T\"{u}t\"{u}nk\"{u}ler H 2002 J. Phys. A \textbf{35} 9425

\bibitem{Ronveaux}Ronveaux A 1995 {\em  Heun's Differential Equations} (Oxford University Press, Oxford, New York)

\bibitem{Slavyanov}Slavyanov S Y and Lay W 2000 {\em  Special Functions: A Unified Theory Based on Singularities} (Oxford University Press, Oxford, New York)

\bibitem{Hu}Hu H and Chen S 2013 arXiv:1302.5933

\bibitem{Forster}F\"{o}rster L, Karski M, Choi J M, Steffen A, Alt W, Meschede D, Widera A, Montano E, Lee J H, Rakreungdet W and Jessen P S 2009 Phys. Rev. Lett. \textbf{103} 233001

\bibitem{Palyi}P\'alyi A, Struck P R, Rudner M, Flensberg K and Burkard G 2012 Phys. Rev. Lett. \textbf{108} 206811

\end{thebibliography}
\end{document}